**Title of the article:**
Commitment to Software Process improvement – Development of Diagnostic Tool to Facilitate Improvement1

**Authors:**
Pekka Abrahamsson

**Notes:**

• This is the author's version of the work.

• Copyright owner's version can be accessed at

https://link.springer.com/article/10.1023/A:1008978919720

# Commitment to Software Process improvement – Development of Diagnostic Tool to Facilitate Improvement[1]

Pekka Abrahamsson

University of Oulu, Department of Information Processing Science
P.O. Box 3000, FIN-90401 Oulu, Finland

**Abstract**

This paper suggests that by operationalizing the concept of commitment in the shape of a model, a new insight is provided in improving software processes - a more human centered approach as opposed to various technical approaches available. In doing so the SPI managers/change agents are able to plan better the software process improvement initiative and benchmark successful projects (as well as failed ones). Results from five interviews with SPI professionals on the proposed Behavior-based Commitment Model are reported, together with early results from the empirical test in 14 software process improvement projects. Early results suggest that the behaviors introduced in the model are relevant in SPI initiatives, the use of model raises the awareness about the people issues in improving processes, and the model could be used aside with CMM, SPICE or other process improvement models.

## 1  Introduction

Commitment has been one of the most popular research subjects in industrial psychology and organizational behavior over the past 30 years (Benkhoff, 1997). It has been believed that successful software process improvement (SPI) depends on the commitment to the project of both managerial levels and software developers (Humphrey, 1989). Indeed, the importance of commitment has been emphasized in the software process community both in the literature (e.g. Humphrey, 1989; Grady, 1997) and in articles for example concerned with the risks that can impede SPI initiatives (e.g. Wiegers, 1998; Statz et al., 1997). Even though the

---

significance of commitment is generally accepted (mainly because of its assumed impact on performance) there has been little research efforts in process improvement field to explore the theoretical foundations and implications to practice from the viewpoint on commitment adopted by the SPI community.

This paper concentrates on reporting early results from an ongoing study aimed at constructing an operational model of commitment (Behavior-based Commitment Model) based on the definition adopted by the software process improvement community. The model consists of two distinct parts: the questionnaire and the framework. The model assumes that the behavior affects the attitude rather than the opposite following the guidelines proposed in self-perception theory explored briefly in the early part of the paper. The Behavior-based Commitment Questionnaire contains nine categories of behaviors identified by Porras and Hoffer (1986) having interviewed leaders in the organizational change field. These nine categories are further divided into individual behaviors. The framework is used to help the user of the model to interpret and communicate the results.

The paper suggests that behaviors, identified in successful organization development efforts by Porras & Hoffer, are relevant in software process improvement initiatives, and the explicit demonstration/consideration of these behaviors by the change agent reflects commitment to the SPI-project and will have a positive effect to the SPI-project's outcome. The Behavior-based Commitment Model was evaluated by conducting five semi-structured interviews with SPI professionals who all had a strong background in improving software processes. All professionals interviewed had a positive attitude toward the model proposed and were interested in testing the model in their projects. Early results from the empirical tests with 14 software process improvement initiatives indicate that all proposed behavior categories have an effect to a SPI-projects and the use of model raises the awareness of the change agents about the people issues in improving processes.

The paper is organized so that it starts of by defining the concept of commitment and by providing a brief view on the underlying theory. The following sections introduce the construction of the Behavior-based Commitment Model, the results from the interviews and the early results from the empirical test. The paper is concluded with a summary and implications of the study to SPI field.

## 2 Background

### 2.1 Defining Commitment

Brown (1996) suggests that for research purposes, useful commitment measures must have a focus, they must specify terms, and they must include a sense of pledge or obligation. He clarifies this by suggesting that all commitments, regardless of the context, share three common components: focus, terms and strength. All commitments have an object or focus – a party to which the commitment is made. Terms are a fundamental part of any commitment since they state the conditions that will fulfill the commitment. Related to the terms, according to Brown, is the strength of a commitment, which will differ depending on the significance or importance in the life of the person who owns the commitment relative to other commitments.

In this paper the focus of commitment is an SPI-project with a clearly defined (as opposed to vaguely defined) and well understood goal(s). It has been well established in the literature that when the goal is too vaguely defined or out of the individual's scope it becomes too abstract to consider (e.g. Robbins, 1993). Gilb (1988) emphasized this point by stating that "*projects without clear goals will not achieve their goals clearly*".

The terms that will fulfill the commitment in this paper are the behaviors that were identified to occur typically in successful organizational changes by Porras and Hoffer [6] since it is commonly accepted that software process improvement will lead to organizational change. This view on commitment is in accordance with a view adopted by the SPI community mainly from Humphrey (1989). Humphrey sites Salancik (1982) when he defines commitment as "*a way to sustain action in the face of difficulties*". This view represents a 'behavioral' school of commitment research (as opposed to an 'attitudinal' school, see e.g. Brown (1996) for details and Abrahamsson (1999a) for an application of such approach in relation to SPI). According to this view, changes in attitude are assumed to be the consequences of changes in behavior, rather than the reverse (Taylor, 1995). Becker (1960) introduced viewing commitment from a behavior oriented point of view as he argued that commitment encompasses structural conditions that make a behavior irrevocable or difficult to change. Later Kiesler (1971) and Salancik (1982) continued exploring this approach.

Strength of a commitment varies depending on the factors fostering commitment. These factors are according to Salancik (1982) publicity, irrevocability, visibility and volitionality of the behavior demonstrated.

Therefore a definition of a measurable commitment that satisfies criteria proposed by Brown (1996) can be defined as a level to which a person explicitly demonstrates his/her commitment by his/her behavior or intended behavior toward SPI-project and stakeholders involved within the project. This definition is further explained by Salancik (1982) as he forms that commitment is a state of being in which an individual becomes bound by his actions and through these actions to beliefs that sustain the activities and his own involvement. Oliver (1990) supports this view as he argues that *it is virtually impossible to describe commitment in any terms other than one's inclination to act in a given way towards a particular commitment target.*

## 2.2 Underlying Theory

The general class of theories in social psychology is known as consistency theories. These theories posit that when attitudes and behaviors are incongruent (within a single individual) the individual will alter either attitudes or behavior to make the two consistent (Menard and Huizinga, 1994). Bem's (1967) self-perception theory asserts that attitudes are used to make sense out of an action (behavior) that has already occurred (Robbins, 1993). Attitude change therefore occurs after the behavior change. From this perspective, attitude is an effect rather than a cause of the behavior (Bem, 1967). Attitudes are also private, non-explicit, and retractable which, according to Salancik (1982), lessens the binding effect that the behavior has on an individual. This theory supports the view on commitment proposed in the paper that when a behavior demonstrated by an individual is volitional, public, explicit and perhaps even nonretractable, the more committing it is and the attitudes will eventually change to be congruent with the behavior.

This paper does not claim that self-perception theory is the only theory underlying the Behavior-based Commitment Model but it provides a well researched corner stone for the model to start with.

Having constructed the literary definition of commitment and introduced the underlying theory, the following chapters will concentrate on operationalizing this definition first by introducing an instrument (Behavior-based Commitment Questionnaire) that can be used to capture the level of commitment[2] and secondly by introducing a framework (Behavior-based Commitment Framework) that allows to interpret and communicate the results.

## 3 Development of the Behavior-based Commitment Model

### 3.1 Instrument to Measure Commitment

Porras and Hoffer (1986) argued that even though much has been written about need for a change process to affect behavior relatively few studies have actually measured behavioral change. They identified common behavior changes in successful organization development effort by interviewing 41 leaders in the field of planned organizational change. It was further argued in the article that identifying a common set of behaviors could provide the basis for measurement instruments that could be used across organizations, thus facilitating the comparison of change efforts. They concluded that 40 out of 41 experts did agree that such common behaviors do exist and that even those who were more used to thinking of change in terms of changed values or beliefs had little trouble identifying typical behavior changes. As a result Porras and Hoffer categorized the behavior changes in all organizational levels to nine categories (later to be referred as the components of the framework) which are defined in Table 1.

**Table 1. Common behavioral changes in successful organization development efforts: Category definitions. Adapted from Porras and Hoffer (1986 p. 485).**

---

[2] Level of commitment in terms of behaviors intended to be demonstrated in the SPI-project.

|    | Behavior category | Description |
|----|---|---|
| C1 | Communicating openly | Behaviors promoting or reflecting the direct giving and receiving of information relevant to getting the process improvement initiative done |
| C2 | Collaborating | Behaviors promoting or reflecting the involvement of relevant persons in the processes of identifying and solving problems. |
| C3 | Taking responsibility | Behaviors reflecting acceptance of responsibility and taking initiative in carrying out process improvement related tasks. |
| C4 | Maintaining a shared vision | Behaviors reflecting a clear formulation, understanding, and commitment to organizational philosophy, values, and purposes and a commitment to high standards. |
| C5 | Solving problems effectively | Behaviors reflecting a problem-solving orientation to difficult prcoess improvement related issues. |
| C6 | Respecting/ supporting | Behaviors reflecting demonstration of respect and support for others as worthwhile individuals. |
| C7 | Facilitating interactions | Behaviors reflecting attention to and use of human process issues in one-on-one, group, and intergroup situations. |
| C8 | Inquiring | Behaviors reflecting a probing, inquiring, diagnostic orientation to the organization and its environment. |
| C9 | Experimenting | Behaviors promoting or reflecting an openness to trying new things. |

Each category shown in Table 1 is further divided to a set of behaviors that Porras and Hoffer identified from the interviews. An example of such division can be seen in Figure 1.

The object of the measurement instrument (Figure 1) is to measure to what extent the change agents[3] (primary users of the model) are going to demonstrate their commitment by taking these behaviors (by behavior category) into consideration and explicitly demonstrating them in their software process improvement projects.

---

[3] The term change agent used in the paper refers to a person or a group of persons who are facilitating and/or responsible for planning and/or coordinating the SPI initiative.

## Figure 1. Behavior-based Commitment Questionnaire

**To what extent do you plan to demonstrate the following behaviors in your upcoming SPI-project?**

Labels in the figure: *component (category) number*, *behavior category*, *scale*, *behaviors*, *selected values*.

**C3 — Taking responsibility:** "Behaviors reflecting acceptance of responsibility and taking initiative in carrying out process improvement related tasks"

Scale columns:
- The behavior is relevant but I do not intent to demonstrate it (0)
- I will demonstrate the behavior to low extent (1)
- I will demonstrate the behavior to moderate extent (2)
- I will demonstrate the behavior to high extent (3)
- The behavior is not relevant in the SPI-project (-)

| ID | Behavior | 0 | 1 | 2 | 3 | - |
|---|---|---|---|---|---|---|
| C3B1 | Figuring out for oneself what is necessary to be effective in one's job and taking initiative for getting whatever information, cooperation, services, or materials are needed from relevant parties inside or outside of the organization. | 0 | 1 | 2 | **3** | - |
| C3B2 | Asking for and taking responsibility and authority. | 0 | **1** | 2 | 3 | - |
| C3B3 | Persisting in the struggle to make needed changes, especially in the face of frustration and ambiguity. | 0 | 1 | **2** | 3 | - |
| C3B4 | Forming and offering more suggestions. | 0 | **1** | 2 | 3 | - |
| C3B5 | Stating one's own contribution to a problematic situation rather than blaming others. | **0** | 1 | 2 | 3 | - |
| C3B6 | Exhibiting behaviors that demonstrate movement along a continuum from monitoring one's own work to managing and prioritizing it to affecting the design of it to affecting its organizational context (e.g., policies and procedures) to affecting the goals and directions of the organization itself. | 0 | **1** | 2 | 3 | - |
| C3B7 | Reflecting the responsibility in expressions of interest and excitement in the work. | 0 | 1 | 2 | **3** | - |
| C3B8 | Reflecting the responsibility in decreased approval seeking, face saving, indifference, burnout, or "coasting". | 0 | **1** | 2 | 3 | - |

General formula:
$$Cx\ \%\ \text{part} = ((CxB1+\ldots+CxBn)/n*3)*100$$

Applying values shown in the example above:
$$C3\ \%\ \text{part} = ((C3B1 + C3B2 + C3B3 + C3B4 + C3B5 + C3B6 + C3B7 + C3B8) / (n * 3)) * 100\%$$
$$= ((3 + 1 + 2 + 1 + 0 + 1 + 3 + 1)/(8*3))*100 = (12 / 24) * 100\%$$
$$= 50.0\%$$

← Value for the C3 that can be ticked to the framework

Figure 1. Behavior-based Commitment Questionnaire[4]

The Behavior-based Commitment Questionnaire consists of the behavior category number (C3 in Figure 1), the behavior category name ('Taking responsibility' in Figure 1) and a set of behaviors that form the category (8 behaviors in Figure 1). Each behavior is evaluated by a change agent as to what extent the particular change agent will demonstrate the behavior under consideration. The scale consists of five points: a) the change agent views the behavior to be relevant, but has no intention to demonstrate it, b) the change agent will demonstrate the behavior to low extent, c) the change agent will demonstrate the behavior to moderate extent, d) the change agent will demonstrate the behavior to high extent, or e) the behavior is not relevant to the SPI-project under evaluation.

The score for each behavior category is calculated separately (example shown in Figure 1) by applying the formula:

$$Cx\ \% = ((CxB1+\ldots+CxBn)/n*3)*100$$

---
[4] Note that the Behavior-based Commitment Questionnaire consists of nine parts, one for each behavior category. Example shown in this figure is the questionnaire for category 3 (Taking responsibility).

In the formula above Cx refers to the component to be calculated, B1 refers to the first behavior, Bn to the last behavior in each category and n is the number of behaviors included in the category (when calculating one should exclude those that were selected as not applicable to the SPI initiative under evaluation). The overall score (component R in Figure 2) over all categories is calculated applying the formula:

$$R \% = ((C1B1+\ldots+C1Bn)+\ldots+(C9B1+..+C9Bn))/(C1n+\ldots+C9n)*3)*100$$

The overall score represents the proportion in percentage of the full potential that will be used to implement software process improvement initiative.

### 3.2 Framework for Interpreting Results

The purpose of the framework is to provide a platform for interpreting and communicating the results obtained from filling out the questionnaire shown in Figure 1. The Behavior-based Commitment Framework (Figure 2) consists of nine behavior categories (categories were defined in Table 1 and represented in Figure 2 as components C1…C9) and the explaining factor as an answer to what the result indicates in relation to the SPI-project.

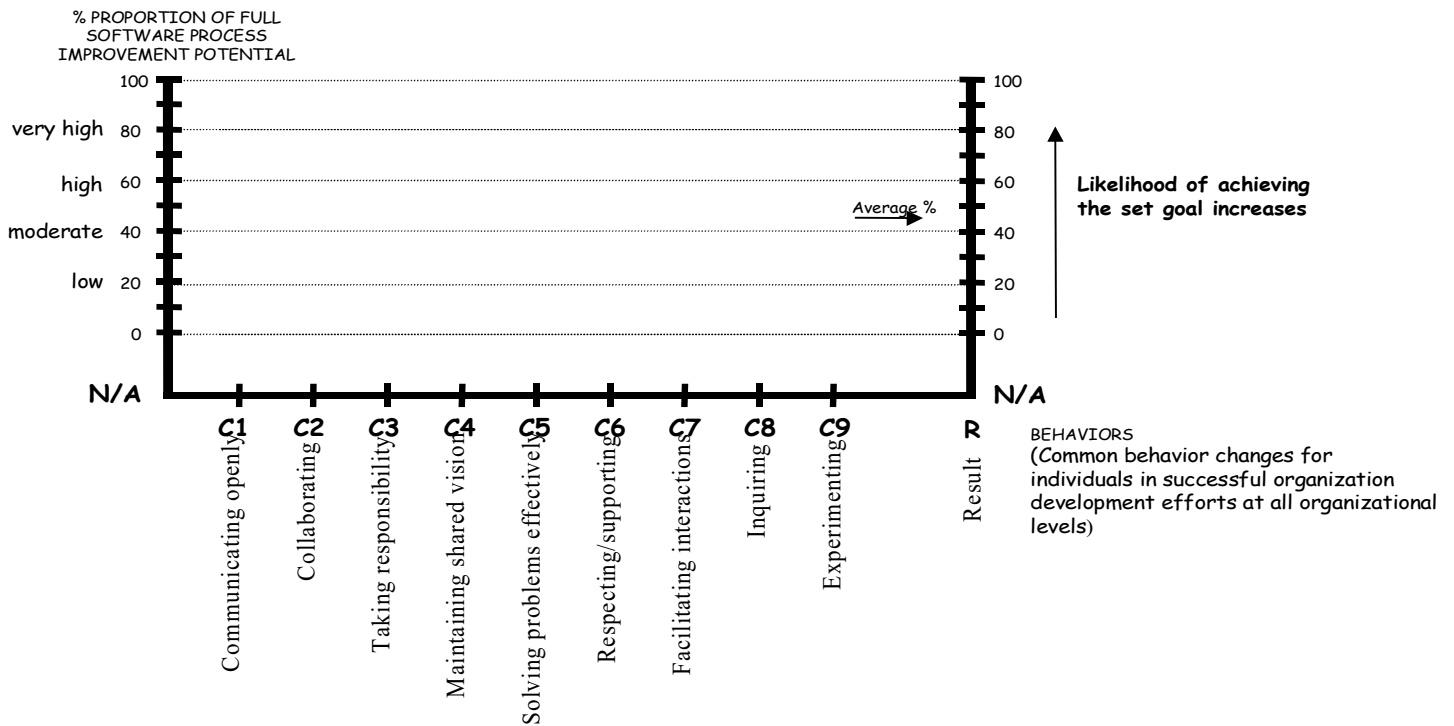

Figure 2. Behavior-based Commitment Framework

The framework suggests (at present state) that all behavior categories and behaviors in those categories are equally evaluated. The author does not claim that though. There is not enough sufficient evidence to weigh certain behaviors over others until the results from the field experiments are thoroughly explored and evaluated.

The underlying hypothesis of the Behavior-based Commitment Model (the questionnaire and the framework together) can be formulated as follows:

1. Behavior changes proposed by Porras and Hoffer (1986) are relevant also in SPI initiatives and

2. the demonstration of these behaviors will have a positive effect to the outcome of the SPI-project.

### 3.3 Applying the Behavior-based Commitment Model in Practice

One should note that when applying the model in practice problems could arise if one takes these behaviors too literally, without considering their context and the underlying values they represent and reflect. In an extreme case, attempting to manipulate individuals to behave in a way identified by the model could arrest individual growth and, in the long run at least, harm the organizational performance as well (Porras and Hoffer, 1986).

The purpose of the behavior-based model is to serve as a medium for designing and carrying through the SPI initiative, and after the project to serve as a diagnostic tool to identify characteristics of a successful/unsuccessful SPI-project.

It is suggested that the SPI-manager together with the SPI-staff fills out the 9-part questionnaire before initiating the SPI-project, calculates the results and uses these results as a basis for a discussion forum with other SPI-managers, SPI-staff and/or software developers (objects of change). When the scores are analyzed, one should keep in mind that the scores do not relate to any absolute values but rather to rough sketches that helps to put the person (or a group) who applies the model to the map. When viewing the model in this manner the scores could be seen as a profile of the SPI initiative under evaluation. If, for example, the SPI-manager together with SPI-staff ranks all behaviors equally high, this indicates that all (high scoring) behaviors are going to demonstrated in a highly visible manner. One should not concentrate on the scores itself but rather on the differences between categories and behaviors to discover which areas should be in the focus of interest. Another suggested possibility for usage is the 'checklist' –format as was pointed out in the interviews (summarized results from the interviews are presented in the following section):

> "[…] it (the Behavior-based Commitment Questionnaire) could be used as a checklist to see if we have covered all the angles. If we say somewhere that 'no' or N/A (not applicable) [for a certain behavior] then we should have a good reason for doing so […]"

After the SPI-project or for example every six months software developers either as individuals (results are calculated using average values) or as a group fills out the Behavior-based Commitment Questionnaire (Figure 1, wording of the question and the scale should be modified to correspond the situation). In this latter situation the purpose is to identify to what extent the software developers feel that these behaviors have been demonstrated in the SPI-project. In this case the results are used to benchmark the SPI-projects.

This paper suggests that the use of the Behavior-based Commitment Model raises the awareness of people applying the model as to what are the positive behaviors that they should consider. Skilling (1996) clarifies this by pointing out that *if one wants to see others change, change one's self.* Demonstrating new, changed behavior works as an indicator to others of the change process' results. By raising the awareness the model brings the human perspective closer to the software process improvement initiatives and works as a forum for discussion and a signal between various stakeholders involved within the project.

## 4    Early Results from Field Experiments
### 4.1    SPI Professionals' Opinions

The Behavior-based Commitment Model was evaluated qualitatively by conducting five semi-structured interviews. All persons interviewed had a strong experience in leading several software process improvement projects. The purpose of the interview was to find out a) if the professionals view commitment from behavior oriented point of view, b) whether the Behavior-based Commitment Model is relevant to SPI-projects, c) where it could be used, and d) how willing are the professionals to try out the model in practice.

The professionals in SPI field characterize commitment using terminology like 'having a strong will', 'doing what we say we do', 'acting first' and 'doing independently'. All respondents agreed that it is one's behavior that ultimately demonstrates one's commitment to improving processes. This supports the behavior-based commitment thinking proposed in the paper.

The results[5] of the interviews suggest that that the model brings extra value to implementing SPI initiatives since professionals thought that the human perspective is rather limited in current process improvement models (e.g. Capability Maturity Model[6], SPICE[7], etc.). All behavior categories were estimated to be at least moderately relevant in SPI-projects (categories 'Open communication' and 'Taking responsibility' were ranked as the highest in relevancy and 'Solving problems effectively' was ranked as the lowest). Professionals viewed several potential usage possibilities for the Behavior-based Commitment Model (see Table 2 for summary). Professionals agreed that the Behavior-based Commitment Model could be used aside with another more technical approach (e.g. CMM). The model proved to be intuitively appealing to the professionals as they all agreed to try out the model in practice.

---

[5] Detailed extracts from the interviews to support the results reported here can be found in the earlier version of the paper (Abrahamsson, 1999b).

[6] Capability Maturity Model and CMM are service marks of Carnegie Mellon University

[7] International Standards Organization's ISO/IEC 15504 standard for Software Process Improvement and Capability dEtermination, formerly known as the SPICE model

This was seen to be an encouraging result from the interview.

The following table (Table 2) summarizes the discussion about the usage possibilities of the model proposed.

| Who | Change agent(s) | Objects of change | Managers |
|---|---|---|---|
| When | Before the project | After the project | Any point in time |
| With whom | SPI group as a whole | Individually or as a group | Individually or as a group |
| Purpose | Plan the project | Evaluate the project | Evaluate the organization |
| Goal | • Raise awareness<br>• Serve as a checkpoint | • Serve as a discussion forum<br>• Diagnose/ benchmark projects | • Raise awareness<br>• Serve as a discussion forum |

Table 2. Behavior-based Commitment Model – usage possibilities

### 4.2 Early Results from the Empirical Test

The Behavior-based Commitment Model is currently been applied in industry in 14 process improvement initiatives. The purpose of the test is to evaluate whether the model works as a diagnostic tool for the change agents and potentially for the software developers and managers as well, fulfilling the purpose and achieving goals proposed in Table 2.

The model was applied by completing the Behavior-based Commitment Questionnaire (an example shown in Figure 1). The questionnaire had two parts for each behavior to be evaluated. In the part A the change agent(s) evaluated whether the behavior in question would affect an SPI-project in general, and in the part B the change agents considered whether the behavior is relevant in their respective SPI-projects and to what extent it should be considered to be demonstrated. Results from the effectiveness of the behaviors to an SPI-project are shown in Figure 3.

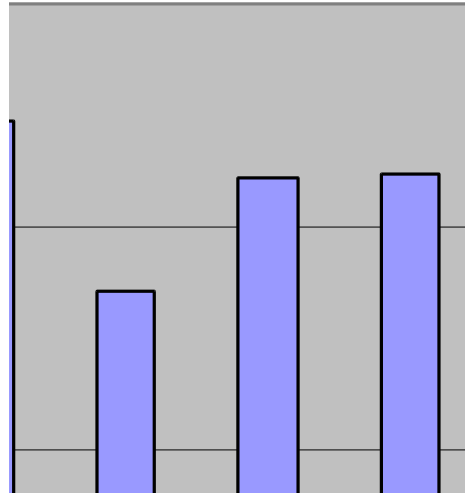

Figure 3. The degree to which a behavior (averaged by categories) affects an SPI-project

Results show that six out of nine behavior categories were evaluated having at least a moderate effect to an SPI-project. These results indicate that the issues pointed out in the model are relevant in conducting an SPI-initiative. The results from the part B as the level to which the change agent(s) plan to demonstrate the behaviors vs. how they evaluated these behaviors to affect the SPI-project is shown in Figure 4[8].

---

[8] In the Figure 4 only behaviors that were relevant to change agents' SPI-projects were used in the calculations. This explains the difference between the effectiveness scores in Figure 3.

Figure 4. The degree to which the behavior affects an SPI-project vs. the degree to which the change agent will demonstrate it.

The results show that the change agent(s) will demonstrate behaviors that were evaluated to have a rather high impact on their SPI-projects and three categories, mainly behavior categories C1, C3, and C6 would be demonstrated the most. Behavior category C4 (maintaining shared vision) was seen to be relevant but will not be demonstrated since the change agents felt that the proposed behaviors were in too abstract level in order to be actually demonstrated.

The resulting (the score for component R in Behavior-based Commitment Framework) score over all categories in all of the projects is shown in figure 5.

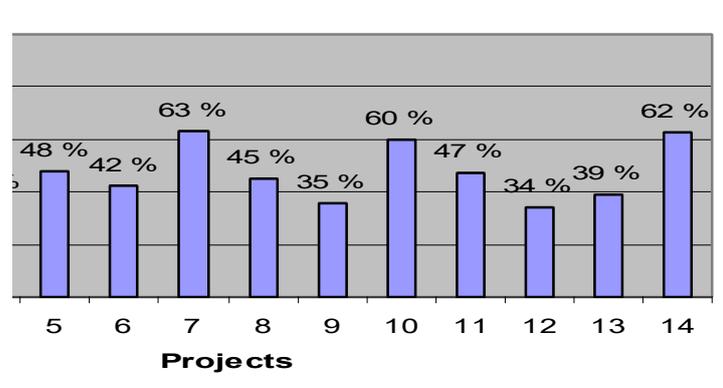

Figure 5. The degree (resulting score) to which the change agent(s) will demonstrate behaviors in their SPI-projects

The author has introduced the results to the change agents and suggested that they regularly evaluate current progress to see whether they really demonstrate behaviors as they have planned. In order to facilitate the follow-up procedure the author highlighted a top ten list of the most important behaviors (based on respective questionnaire) for each SPI group to monitor. This was seen to be a useful approach to communicate the results.

Early findings suggest that this model does raise awareness of the change agents about the people issues in improving processes. Another finding was that the Behavior-based Commitment Questionnaire contains currently too many behaviors (72 behaviors currently) to be evaluated. Some of the behaviors were also found to be too abstract for change agents to consider (especially in 'maintaining shared vision' category). Whether raising awareness and an active demonstration of the behaviors will enhance the likelihood of achieving the goals remains to be proven. All software process improvement projects participating the study are going to be finished by the end of 1999.

## 5    Conclusion

This paper has described early results of an ongoing study aimed at constructing an operational model of commitment (Behavior-based Commitment Model) based on the view on commitment adopted by the software process improvement community. The model along the underlying theory and the operationalized definition was introduced with suggestions on how and where to apply it.

It has been suggested that by definition, in the software process improvement field, behaviors are thought to affect attitudes of an individual. The supporting theory is Bem's (1967) self-perception theory that argues that attitude change occurs after the behavior change rather than the reverse. The paper argued further that the demonstration of behaviors that have been identified occurring in successful organizational changes not only reflects commitment to SPI-project but will also have a positive effect on the outcome of the SPI-project.

The proposed model was evaluated by interviewing five SPI professionals and is currently been applied in 14 process improvement initiatives. Early findings from the empirical test and the fact that the professionals agreed on the relevancy of the behavior categories proposed by Porras and Hoffer (1986) in the SPI-projects supports the first hypothesis proposed in the paper. The second hypothesis, whether the use of the model by change agents will indeed have a positive effect to the outcome of the project, needs to be proven. Results in near future will show what are the benefits and challenges of the use of the Behavior-based Commitment Model.

The positive feedback received from the professionals in the interviews demonstrated that there is a need in the SPI community to include a human element in the models that guide the improvement process. This is not to say that only the human elements are important but to emphasize its meaning in the long run that SPI field needs to gain better understanding of such complex processes as motivating people and committing them to organizational change as it is the case in improving software processes.

Little research has been conducted in this area of study in the software process improvement field. This research opens views on commitment that have not been explored in software process improvement literature before and by doing this will provide a starting point to a discussion forum for the research specialists also. Future research could focus on identifying recurring behavior patterns that are associated with successful or failing process improvement initiatives. These behavior patterns could be identified as behavioral repertoires[9] or behavioral pools and could be used to increase understanding of the human factors in improving software processes. A solid starting point for identifying such is the Behavior-based Commitment Model proposed in the paper.

---

[9] Dobni et al. (1997) introduced and used the term behavioral repertoires when identifying specific combinations of behaviors that comprised service employee roles.

This paper provides a new insight into viewing commitment not only as a psychological state but also as a diagnostic tool that helps change agents to plan better their improvement projects. One of the professionals interviewed emphasized this point by concluding that:

> "It can be said that here is all you need to do. If you do not have this (Behavior-based Commitment Questionnaire) in hand, you would remember only 5% from it. With the paper you would remember 25% […] The advantage of this (Behavior-based Commitment Model) is that it gives you focus, and could work as a discussion forum […] This sounds to me very valuable."

**References**

_